\documentclass[twocolumn,superscriptaddress,amsmath,amssymb,showpacs]{revtex4}

\usepackage{graphicx}
\usepackage{epsfig}
\usepackage{hyperref}
\usepackage{dcolumn}
\usepackage{amssymb}
\usepackage{amsmath}
\usepackage{natbib}
\usepackage{subfigure}
\usepackage{color}

\begin{document}
\title{Noisy Quantum Teleportation: \\An Experimental Study on the Influence of Local
Environments}
\author{Laura T. \surname{Knoll}}
\affiliation{DEILAP, CITEDEF, J.B. de La Salle 4397, 1603 Villa Martelli,
Buenos Aires, Argentina}

\author{Christian T. \surname{Schmiegelow}}
\affiliation{DEILAP, CITEDEF, J.B. de La Salle 4397, 1603 Villa Martelli,
Buenos Aires, Argentina}

\author{Miguel A. \surname{Larotonda}}
\affiliation{DEILAP, CITEDEF, J.B. de La Salle 4397, 1603 Villa Martelli,
Buenos Aires, Argentina}

\begin{abstract}
We report experimental results on the action of selected local environments on
the fidelity of the quantum teleportation protocol, taking into account
non-ideal, realistic entangled resources. Different working conditions are
theoretically identified, where a noisy protocol can be made almost insensitive
to further addition of noise. We put to test these conditions on a photonic
implementation of the quantum teleportation algorithm, where two polarization
entangled qubits act as the entangled resource and a path qubit on Alice encodes
the state to be teleported. Bob's path qubit is used to implement a local 
environment, while the environment on Alice's qubit is simulated as a weighed
average of different pure states. We obtain a good agreement with the theoretical 
predictions, we experimentally recreate the conditions to obtain a noise-induced 
enhancement of the protocol fidelity, and we identify parameter regions of increased 
insensibility to interactions with specific noisy environments.
\end{abstract}

\maketitle

\section{Introduction}
\label{sec:intro}

In the quantum teleportation algorithm an unknown quantum state is transmitted between two distant parties who share an entangled pair of qubits \cite{bennett1993teleporting}. The unknown quantum state is destroyed by the sender (Alice) and later reconstructed non-locally by the receiver (Bob) using purely classical information and quantum entanglement.
Quantum teleportation relies on entanglement as the quantum resource to be able to communicate with higher fidelities than those that  can be achieved by classical communications. Ideally, in the quantum teleportation protocol, the state can be reconstructed with perfect fidelity $F=1$, while it has been shown that the maximum fidelity attainable by a purely classical channel is $F=2/3$ \cite{massar1995optimal}.
Nevertheless, the entangled pair may suffer from decoherence by the interaction with the environment, producing mixed entangled states. Popescu showed that a mixed state could still be useful for teleportation with fidelities greater than the classical limit \cite{popescu1994bell} \cite{bennett1996purification}, although not every mixed entangled state can achieve quantum fidelities \cite{horodecki1996teleportation}.
The fidelity of teleportation can be related to the \emph{fully entangled fraction} $f$ of the quantum resource $\rho$, defined in \cite{bennett1996mixed} as
\begin{equation}
	f(\rho)=\max\limits_{\psi}\langle\psi|\rho|\psi\rangle .
	\label{eq:fef}
\end{equation}

If Alice and Bob share a state $\rho\in\mathcal{H_A}\otimes\mathcal{H_B}=C^{d}\otimes C^{d}$
the teleportation fidelity \emph{F} can be defined in terms of the fully entangled fraction as \cite{horodecki1999general}
\begin{equation}
	F=\frac{fd+1}{d+1}.
	\label{eq:fidelity}
\end{equation}

For the case $d=2$ we obtain $F=\frac{2f+1}{3}$. In order to improve the classical teleportation limit  ($F>2/3$) we need $f>1/2$. Hence the fidelity on a teleportation protocol can be increased by increasing $f$.

Badziag \textit{et. al.} \cite{badziag2000local}
presented a class of two-qubit states with $f<1/2$ which can 
be used for teleportation with non-classical fidelity. They showed that dissipative
interactions with an environment via an amplitude damping channel can enhance
the fidelity of teleportation for a family of non-teleporting mixed entangled
states, allowing for teleportation with quantum fidelity. Bandyopadhyay
\cite{bandyopadhyay2002origin} later analyzed the action of the amplitude damping channel for
bipartite maximally entangled states (the four Bell states) where the qubits of
the entangled pair undergo local interactions with their respective
environments. In \cite{yeo2008local} it was shown that an enhancement in teleportation fidelity via dissipative interactions with the local environment can be achieved for two-qubit teleportation using a family of four-qubit mixed states as a resource. More general conditions to establish an optimal teleportation protocol when noise is present both in the quantum channel and in the input state were presented and studied in \cite{taketani2012optimal}.\\ 

In this work we focus on experimental results from a quantum teleportation scheme that is afflicted by local noise. By evaluating the quantum fidelity of the process we study the influence of particular dissipative interactions, and we analyze different scenarios where the protocol shows different sensitivities to local noise, with the goal of obtaining practical and robust conditions for quantum communications. The following section is devoted to a description of the formal aspects of the problem, and the specific noise channels that are experimentally implemented. Section \ref{sec:dissipenv} describes the action of the local environments on entangled resource and its effect on the fully entangled fraction of the quantum resource. Section \ref{sec:experiment} describes the experimental realization of the noisy teleportation protocol, and finally section \ref{sec:results} is devoted to the presentation and discussion of the results.

\section{Interaction with dissipative environments}
\label{sec:theory}
\subsection{General framework}
The dynamics of open quantum systems can be studied as the interaction of a system $S$ with an environment $E$. We consider a system in state $\rho$, which is transformed into the final state $\varepsilon(\rho)$ by an evolution operator $U$. We also assume that the system and the environment are initially in a product state $\rho\otimes\rho_{E}$: after the evolution the system no longer interacts with the environment, so a partial trace can be performed over the environment to obtain the evolution of system $S$ \cite{nielsen2010quantum}:
\begin{equation}
 \varepsilon(\rho)=Tr_{E}[U(\rho\otimes\rho_E)U^{\dagger}].
\end{equation}

Let $|e_k\rangle$ be an orthonormal basis for the state space of the environment, and let the initial state of the environment be $\rho_{E}=|e_0\rangle\langle e_0|$, without loss of generality. The previous equation can then be rewritten as
\begin{equation}
\begin{aligned}
 \varepsilon(\rho)&=\sum_j\langle e_j|U(\rho\otimes|e_0\rangle\langle e_0|)U^{\dagger}|e_j\rangle \\
 &=\sum_j K_j \rho K_j^{\dagger}
\end{aligned}
 \end{equation}
where $K_j=\langle e_j|U|e_0\rangle$ are the Kraus operators \cite{kraus1983states}. These operators are trace preserving, meaning that $\sum_j K_jK_j^{\dagger}=1$, which guarantees that $Tr[\varepsilon(\rho)]=1$. Throughout this work we implement and study the influence of the amplitude damping and phase damping dissipative channels on the teleportation protocol. The properties of their interactions with the system are described as follows:
\begin{itemize}
 
\item \emph{Amplitude Damping Channel} (ADC), which describes the energy dissipation of a system to the environment. It is typically an atomic process, and it can be thought of as the spontaneous emission of a photon into the environment by the decay of an excited state of a two-level atom in the presence of an electromagnetic field. This channel can be described with the map
\begin{equation}
\begin{aligned}
	&|0\rangle_S|0\rangle_E\rightarrow|0\rangle_S|0\rangle_E\\
	&|1\rangle_S|0\rangle_E\rightarrow\sqrt{1-p}|1\rangle_S|0\rangle_E+\sqrt{p}|0\rangle_S|1\rangle_E.
	\label{eq:ADCmap}
\end{aligned}
\end{equation}

When there is no excitation present, the system and the environment remain unaltered, while when an excitation is present in the system, it can either remain there with probability $(1-p)$ or decay to the ground state with probability $p$, producing an excitation in the environment. This map can be put in the form of Kraus operators as
 \begin{equation}
K_{1}=
\left( \begin{array}{cc}
1 & 0 \\
0 & \sqrt{1-p} \end{array} \right)\quad
K_{2}=
\left( \begin{array}{cc}
0 & \sqrt{p} \\
0 & 0 \end{array} \right).
\label{eq:krausADC}
\end{equation}

\item \emph{Phase Damping Channel} (PDC), in which the system looses coherence due to -for example- random scattering without any loss of energy. This process only involves the phase, and it can be associated to a loss of coherence between photonic states due to a temporal or transversal mismatch. The map that describes this noisy channel is the following:
\begin{equation}
\begin{aligned}
	&|0\rangle_S|0\rangle_E\rightarrow|0\rangle_S|0\rangle_E\\
	&|1\rangle_S|0\rangle_E\rightarrow\sqrt{1-p}|1\rangle_S|0\rangle_E+\sqrt{p}|1\rangle_S|1\rangle_E.
	\label{eq:PDCmap}
\end{aligned}
\end{equation}
and the corresponding Kraus operators are:
 \begin{equation}
K_{1}=
\left( \begin{array}{cc}
1 & 0 \\
0 & \sqrt{1-p} \end{array} \right)\quad
K_{2}=
\left( \begin{array}{cc}
0 & 0 \\
0 & \sqrt{p} \end{array} \right).
\label{eq:krausPDC}
\end{equation}

\end{itemize}

In the following section, we formally describe the interaction of an entangled qubit pair with local environments, and we calculate the fully entangled fraction of such perturbed state, as a measure of its ability to serve as the entangled resource for quantum teleportation.

\subsection{Effect on a two-qubit entangled state}
\label{sec:dissipenv}
We shall consider different cases, where the entangled resource available for the teleportation protocol is initially a two-qubit pure Bell state
\begin{equation}
  |\Phi^{+}\rangle=\frac{1}{\sqrt{2}}\left(|00\rangle +|11\rangle\right),
\end{equation}	
that interacts with the environment via an amplitude and/or a phase damping channel.\\

For the case where only one qubit from the maximally entangled state undergoes dissipation via ADC, suppose Alice's qubit interacts with the environment. Using the
Kraus operators formalism, the evolution of state $\rho$ into $\rho'$ may be described as
\begin{equation}
  \rho'=\varepsilon(\rho)=A_{1}\rho A_{1}^{\dag}+A_{2}\rho A_{2}^{\dag}
\label{eq:mapa_epsilon}
\end{equation}
where $A_{i}=K_{i}\otimes I$ and $K_i$ are the Kraus operators  for the ADC map \eqref{eq:krausADC}.

The state $\rho'$ may then be written as
\begin{equation}
  \rho'=\frac{1}{2}
  \left( \begin{array}{cccc}
    1 & 0 & 0 & \sqrt{1-p_{a}}\\
    0 & p_{a} & 0 & 0 \\
    0 & 0 & 0 & 0 \\
    \sqrt{1-p_{a}} & 0 & 0 & (1-p_{a})
    \end{array} \right).
    \label{eq:ADCalice}
\end{equation}
\\
The fully entangled fraction for this state, defined in Eq.\eqref{eq:fef} and calculated using the closed-form expression from \cite{grondalski2002fully}, is
\begin{equation}
  f_{A}(\rho')=\frac{1}{4}\left(1+\sqrt{1-p_{a}}\right)^2.
\label{eq:ADCsingle_local}
\end{equation}
For the state to be useful for teleportation (i.e. to be able to achieve
teleportation fidelities above the classical limit) the fully entangled fraction
must satisfy $f(\rho')>1/2$. We will refer to states that satisfy this inequality as ``teleporting states''. It is easy to see that
\begin{equation}
  f_{A}(\rho')>\frac{1}{2}\hspace{0.3cm}\Leftrightarrow\hspace{0.3cm} p<2\sqrt{2}-2.
\end{equation}

If we now let Bob's qubit in $\rho'$ interact with an environment via ADC as
well, the transformed state $\rho''$ is now
\begin{equation}
  \rho''=\varepsilon(\rho')=A_{1}\rho' A_{1}^{\dag}+A_{2}\rho' A_{2}^{\dag}
\end{equation}
with operators $A_{i}=I\otimes K_{i}$. In this case, the fully
entangled fraction for the state $\rho''$ after the interaction of both qubits
with an amplitude damping channel, with their respective dissipative parameters
$p_{a}$ for Alice and $p_{b}$ for Bob, is now


\begin{equation}
    f_{AA}(\rho'')=\frac{1}{4}\left[p_{a}p_{b}+\left(1+\sqrt{(1-p_{a})(1-p_{b})}\right)^2\right].
\label{eq:fef_2q}
\end{equation}

\begin{figure}[h!]
    \centering
    \includegraphics[width=0.47\textwidth]{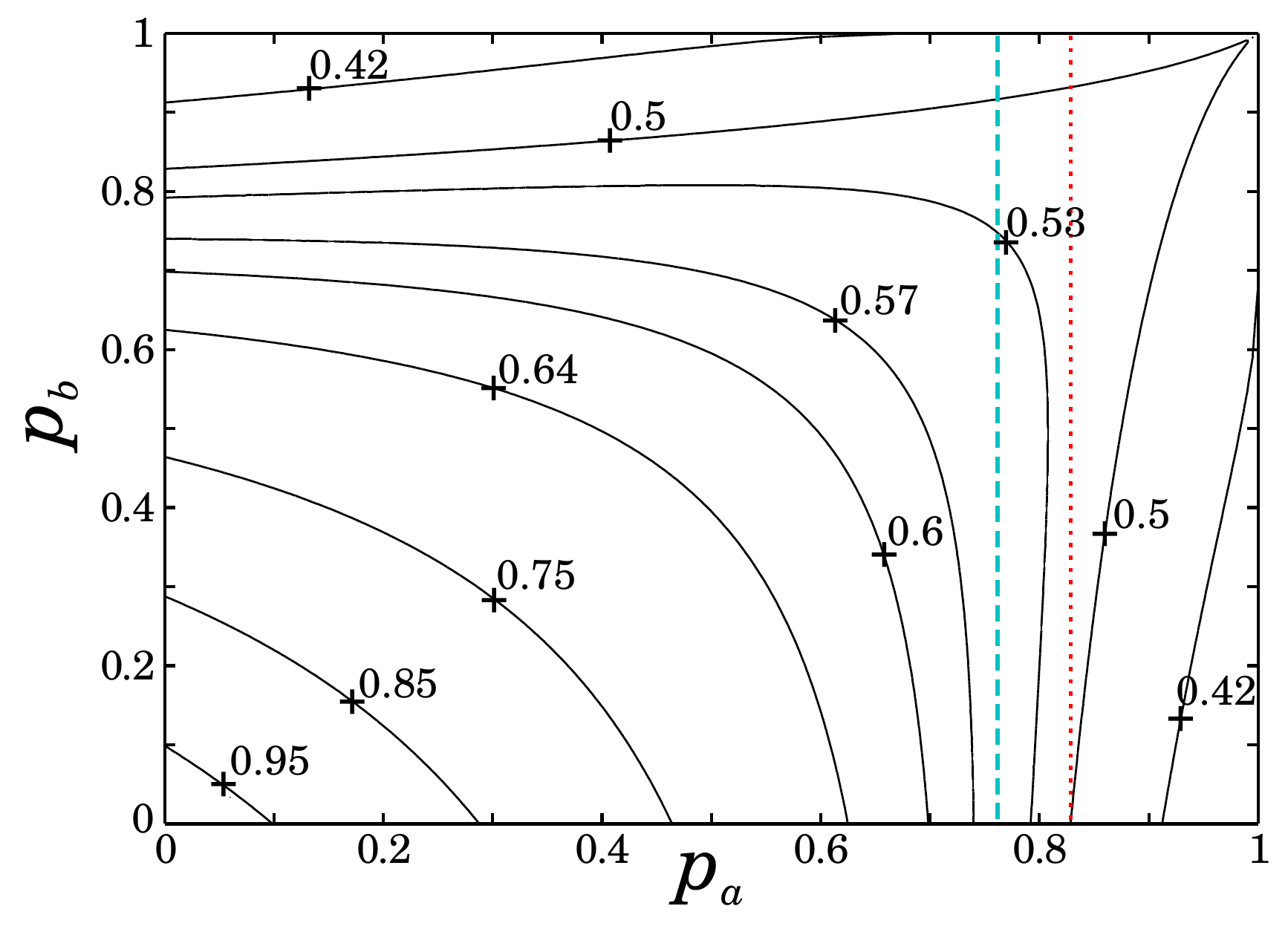}
  \caption{(Color online) Contour plot for the fully entangled fraction as
a function of both damping parameters $p_a$ and $p_b$. Blue dashed line: $p_a\approx$0.76, where the fully entangled fraction is almost constant for damping strengths $p_b$ between 0 and 0.7. Red dotted line: for a constant value of $p_a=2\sqrt{2}-2$, the fully entangled fraction is 0.5 with no dissipation on the other qubit and rises up to 0.52 for $p_b$=0.602}
  \label{fig:fef}
  \end{figure}

It is worth to note that since both subsystems are affected by similar local interactions, the effect on the fully entangled fraction is symmetric with respect to $p_a$ and $p_b$.
We now analyze different scenarios for each damping parameter, recalling that the teleportation fidelity is directly related to the fully entangled fraction of the entangled resource via \eqref{eq:fidelity}:
\begin{itemize}

  \item If only one of the qubits suffers dissipation (let $p_{a}$=0) the fidelity is maximum ($F$=1) for $p_{b}$=0, and monotonically decreases to reach the classical limit $F$=2/3 ($f$=1/2) at $p_b=2\sqrt{2}-2$. Under larger damping conditions the initial maximally entangled resource becomes highly inefficient for teleportation ($\rho''=\frac{1}{2}\left(|00\rangle \langle00| +|01\rangle \langle01|\right)$ at $p_b$=1), hence the fidelity falls below the classical limit, down to a value of 1/2 for $p_b$=1.

  \item If both interactions occur with the same parameter $p=p_{a}=p_{b}$ the
fully entangled fraction may be written as
\begin{equation}
  f(\rho'')=\frac{1}{2}\left[1+\left(1-p\right)^2\right],
\end{equation}
from where we obtain $f(\rho'')>1/2$ $\forall$ $0\leq p<1$. 

Remarkably, even though noisy environments destroy the coherence of entangled pairs, letting both qubits
interact with an environment via ADC with the same damping parameters gives an
enhancement on the achievable teleportation fidelities with respect to the ones
obtained when only one qubit interacts with the environment. In other words, for certain bipartite quantum states used as the teleportation resource, the fidelity of the process can  be enhanced by a local 
dissipative interaction \cite{badziag2000local,bandyopadhyay2002origin}. For $p_b$ \emph{and} $p_a$ equal to $p=2\sqrt{2}-2$, using \eqref{eq:ADCsingle_local} and \eqref{eq:fidelity} it can be seen that the process fidelity raises to the non-classical value 
\begin{equation}
F=2(3-2\sqrt{2})+1/3\approx0.6765;
\end{equation}
in particular, non-teleporting states can be made teleporting by dissipative interactions with
an environment.

  \item The previous scenario imposes some stringent and unrealistic situation, in which both damping strengths are known and at least one can be controlled. Nevertheless, for $p_{a}\neq p_{b}$ we can find different regions in the parameter space where the
fully entangled fraction is greater than 1/2. Fig. \ref{fig:fef} shows a contour
plot for Eq. \eqref{eq:fef_2q} as a function of both damping parameters. If one of the parties can generate such a dissipative interaction, then the fidelity of a teleportation process under local amplitude damping can be enhanced by tuning the damping strength of the controlled environment. 

  \item Also interesting is the fact that for fixed values of $p_{a}$ ($p_{b}$), we can find curves with $f$ values greater
than 1/2 where the influence of noise is almost negligible for all values of
$p_{b}$ ($p_{a}$), as the fully entangled fraction (and hence the fidelity) remains approximately
constant (Fig. \ref{fig:fef}). 

\end{itemize}

\begin{figure}[h!]
    \centering
    \includegraphics[width=0.47\textwidth]{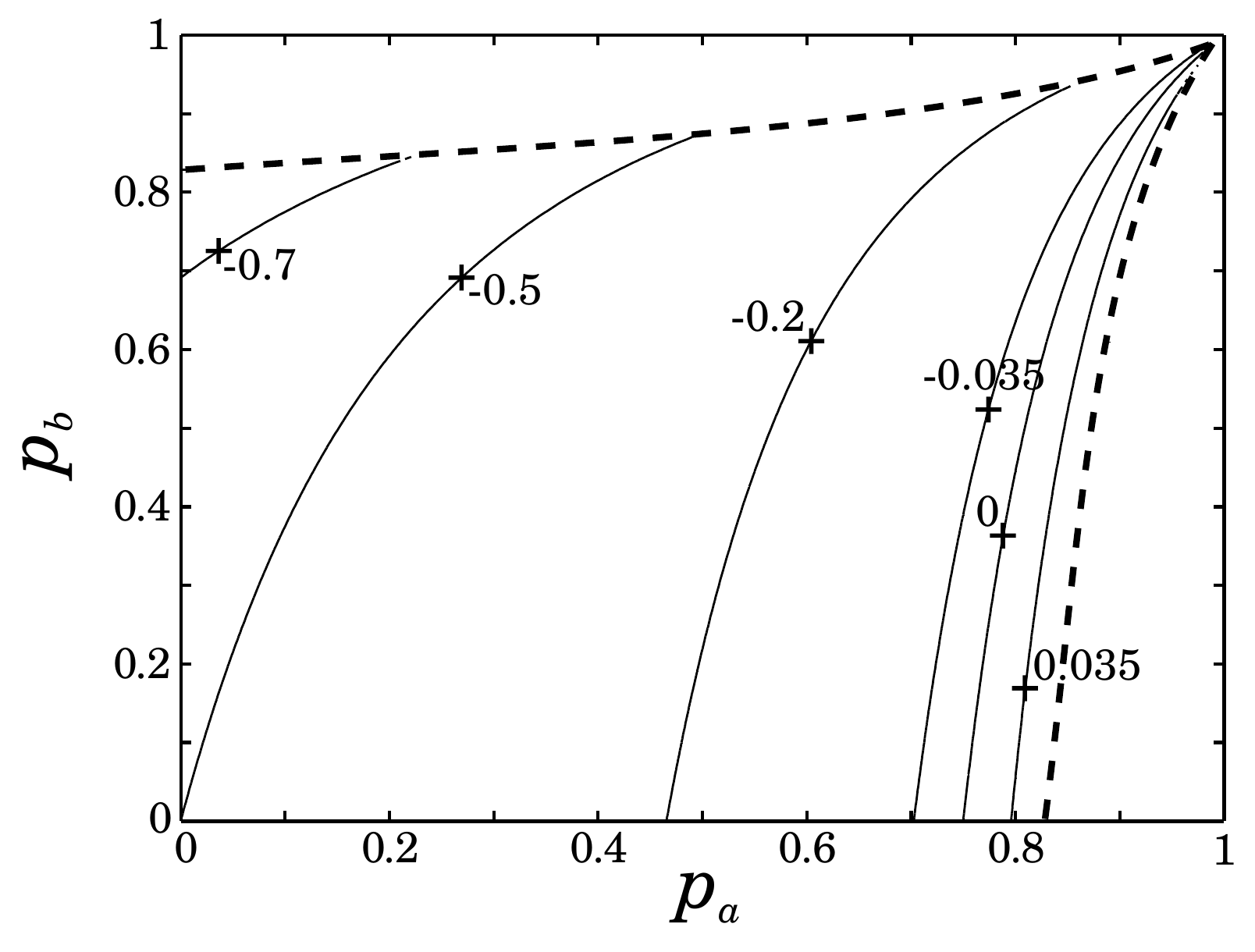}
  \caption{Plot of $\partial f/\partial p_b $ in solid lines, for values of the fully entangled fraction above 1/2, as
a function of both damping parameters $p_a$ and $p_b$. Regions of low sensibility to the coupling strength, and even regions of positive slope (increase of the fidelity with increasing damping) can be observed. The dashed line corresponds to the classical teleportation limit $f$=1/2.}
  \label{fig:diff}
  \end{figure}

We should stress the fact that the above conclusions are based on the assumption that the original teleporting resource is a pure, maximally 
entangled state, and specific local noise is added. A natural question that emerges is how these results are 
affected when the entangled resource is imperfect, and also if the predicted fidelity enhancement is still 
attainable in a real world implementation of the protocol. 
In such cases, gaining knowledge on the sensibility of the teleportation efficiency under the variation of the 
strength of a local operation may also prove to be of practical interest. When working in realistic laboratory 
conditions, it might then be useful to let the qubit undergo controlled dissipative interactions, so as to 
mitigate the sensibility to noise but still be able to teleport with quantum fidelities. Fig. \ref{fig:diff} shows a plot 
of $\partial f/\partial p_b $, for values of the fully entangled fraction above 1/2: in this ideal case, regions where the fully entangled fraction is insensitive to this 
particular noise are present, even under strongly dissipative conditions. In particular there is a solution for $\partial f/\partial p_b $=0, meaning that $f_{AA}(\rho'')$ is not a monotonically decreasing function and therefore an increase of the fidelity can be expected.

Finally we present the expressions for the fully entangled fraction of a two-qubit maximally entangled state that is affected by a phase damping channel on one qubit ($b$) and an amplitude damping channel on the other qubit ($a$);

\begin{equation}
    f_{AD}(\rho'')=\frac{1}{2}\left[1 + \sqrt{(1-p_{a})(1-p_{b})} -\frac{p_{a}}{2}\right],
\label{eq:fef_2qAD}
\end{equation}

and when the dissipative interactions are phase damping channels on both qubits:

\begin{equation}
    f_{DD}(\rho'')=\frac{1}{2}\left[1 + \sqrt{(1-p_{a})(1-p_{b})} \right].
\label{eq:fef_2qDD}
\end{equation}

It is easy to check that the partial derivatives of (\ref{eq:fef_2qAD}) and (\ref{eq:fef_2qDD}) are always negative in the interval $0\leq p_{a},p_{b} \leq 1$ and therefore we cannot expect a fidelity enhancement with an increase of any of the damping strengths. 

In the next section we address these points from the experimental point of view, using a previously 
reported quantum teleportation setup \cite{knoll2014remote} that relies on an inherently imperfect entangled resource,
 where we add a local noisy environment on Bob's qubit and we simulate a local environment on Alice's side by averaging measurements obtained with pure states. All the results are expressed in terms of the process fidelity $F$, which can be directly obtained from the experimental results.

\section{Experiment}
\label{sec:experiment}
\subsection{Teleportation Setup}
\begin{figure}[h!]
    \centering
    \includegraphics[width=0.48\textwidth]{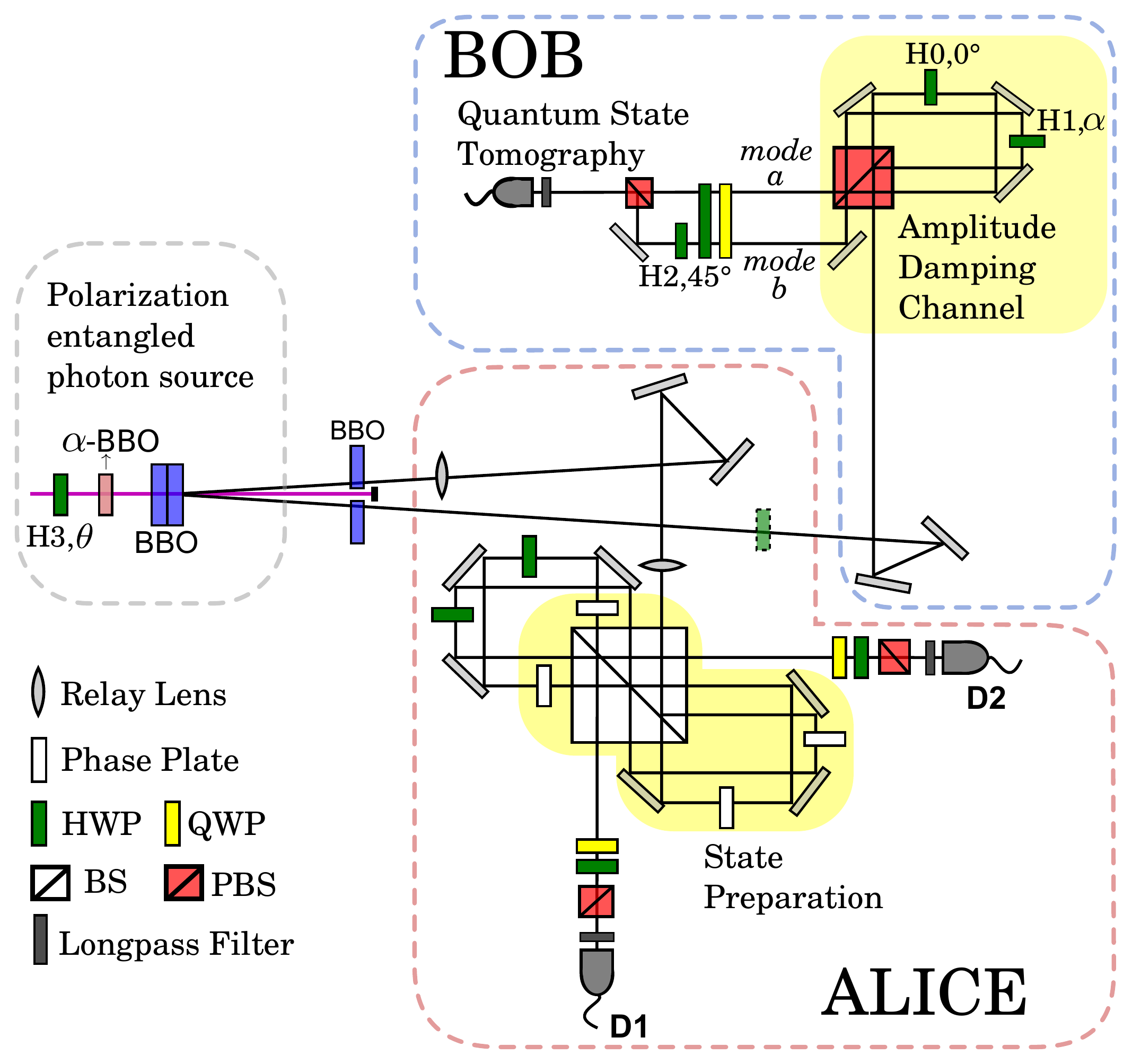}
  \caption{(Color online) Experimental setup for the implementation of the
teleportation algorithm and noisy environments. The entangled SPDC pair source is 
optimized using temporal and spatial compensating birefringent crystals. State
preparation is performed on a path qubit on Alice's side using a displaced
Sagnac interferometer and phase plates. The controlled-not gate is implemented with
half-wave plates placed on each of the logical states of the path qubit, one at
$0^{\circ}$ and the other at $45^{\circ}$, whereas the Hadamard gate is obtained
with a third passage through the beam splitter. Quantum state tomography is
applied on Bob's photons in coincidence with a detection of a photon on one of
Alice's outputs. Bob's polarization qubit can be coupled to a local environment encoded 
on the path qubit (see text): the photon enters a Sagnac interferometer based on a PBS,
where the $H$ and $V$ polarization components are routed in different directions. 
If $\alpha=0^{\circ}$ both polarization components are coherently recombined in the PBS and exit the
interferometer in mode $a$. For every other angle of $\alpha$ the $V$ component
is transformed into an $H$ polarized photon with probability $p=\sin^{2}(2\alpha)$
and exits the interferometer in mode $b$. Both modes are later incoherently recombined
using a HWP oriented at $45^{\circ}$ on mode $b$ (H2) and another polarizing beam splitter, 
which corresponds to a partial tracing operation over the environment. The light green-dashed HWP placed on Bob's path
is occasionally used to generate state $\rho'_2$ to simulate the noisy channel on Alice's side (see text).}
  \label{fig:arreglo}
  \end{figure}

The experimental arrangement that implements quantum state teleportation was originally reported in \cite{knoll2014remote}, and uses a polarization-entangled photon pair source as the quantum resource, while the state to be teleported is encoded in the path of one of the photons. Photon pairs are generated by spontaneous parametric downconversion (SPDC) in a BBO type-I nonlinear crystal arrangement \cite{kwiat1999ultrabright}, pumped by a 405nm CW laser diode. Decoherence originated from timing information and spatial mode phase dependence generated by the birefringence of the nonlinear crystals is corrected using a series of birefringent compensating crystals introduced on the pump beam and on the photon pair paths. The preparation of the path-encoded input state is obtained using a Sagnac displaced interferometer with a variable relative phase between paths and an additional relative phase at the output. The inherent stability of this kind of interferometers allowed us to achieve 98\% visibility at the 
output 
measured in coincidence with Bob's detections. The controlled-not gate ---with the polarization as the target qubit--- is obtained with half waveplates inserted on a similar displaced Sagnac interferometer arrangement, and the Hadamard gate needed for the protocol is implemented on the path qubit using a beam splitter. The two interferometers share the same beam splitter cube, forming a bow-tie layout. The complete setup is depicted in Fig. \ref{fig:arreglo}.

Measurements are obtained by projecting Alice's output onto the canonical two-qubit basis states \{$|0H\rangle$, $|0V\rangle$, $|1H\rangle$, $|1V\rangle$\}, where the first qubit represents the path qubit and the second the polarization qubit.
The teleported state is reconstructed at Bob's side by performing standard quantum state tomography \cite{nielsen2010quantum,james2001measurement} of the polarization state in coincidence with any of the four projective measurements obtained by Alice, and performing the corresponding rotations \cite{bennett1993teleporting}. The preparation of the different path states, the polarization selection of Alice's output and the polarization projections to perform tomographic measurements on Bob's qubit are all automated, using actuators built in our laboratory, based on servo motors and stepper motors.

In the $\chi$-matrix representation, the quantum process for a single qubit is 
described by $\varepsilon(\rho)=\sum\nolimits_{mn}\chi_{mn}E_m\rho E_n^{\dagger}$ \cite{nielsen2010quantum}.
For processes which can be described by a $\chi$-matrix the average fidelity can be 
easily calculated from the $\chi_{00}$ element \cite{renes2004symmetric} as
\begin{equation}
  F_{AV}=\frac{2 \chi_{00}+1}{3}
  \label{eq:avfidel}
\end{equation}
provided that the operator basis
$\left \{E_m \right \}$ satisfies: $Tr\left ( E_mE_n\right )=2\delta_{mn}$,
$E_mE_m^{\dagger}=I$, and $E_0=I$ \cite{bendersky2009selective}. In our case the operator basis is $\left\{I,X,Y,Z\right \}$.
Via quantum process tomography, the $\chi$-matrix was reconstructed from the measurements for the four possible outcomes. The teleportation protocol requires to apply an additional rotation that depends on the measurement of Alice's qubits to recover the orginal state. Therefore, the element $\chi_{00}$ of the complete process can be operationally obtained using the corresponding diagonal element of each of the reconstructed processes $\chi^{(ij)}$ associated with the different outcomes of Alice's qubits: $i=0,1$, $j=H,V$. The final value for the fidelity is then obtained as:
\begin{equation}
 F=\frac{2\left(p_{0H}\chi^{(0H)}_{II}+p_{0V}\chi^{(0V)}_{XX}+p_{1H}\chi^{(1H)}_{YY}+p_{1V}\chi^{(1V)}_{ZZ}\right)+1}{3}
 \label{eq:fidel_proba}
\end{equation}
with $p_{ij}=Tr(P_{ij}\rho_{out}P_{ij}^{\dagger})$, $\rho_{out}$ is the state obtained after applying the teleportation protocol and $P_{ij}$ the projector on the $|ij\rangle$ state outcome. $p_{ij}$ is therefore the probability of finding the teleported state when the $|ij\rangle$ outcome is measured. If the two-qubit state is maximally entangled, ideally $p_{ij}=1/4$ and $\chi^{(0H)}_{II}=\chi^{(0V)}_{XX}=\chi^{(1H)}_{YY}=\chi^{(1V)}_{ZZ}=1$.


\subsection{Noisy environments}

\emph{Local noise on Bob's qubit.} For the implementation of the amplitude damping channel on Bob's side we used
the arrangement proposed by the group of S. Walborn at UFRJ in \cite{almeida2007environment}. Fig.\ref{fig:arreglo} shows a schematic representation 
of the Rio environment implementation on Bob's side. In this scheme, Bob's polarization qubit interacts
with a controlled environment encoded in the path qubit, using a displaced Sagnac interferometer based on a
polarizing beam splitter (PBS). If we associate the horizontal ($H$) polarization component with
the ground state and the vertical polarization ($V$) with an excited state, and the environment is
represented by the two path modes of the photon, map \eqref{eq:ADCmap} is
implemented as follows: an incoming photon on mode $a$ is split into its $H$ and
$V$ components by the PBS. The $V$ polarized photons are reflected and propagate
inside the interferometer in the counter-clockwise direction, passing through a
half-wave plate (H1) which transforms the vertical polarization state to a
$\cos(2\alpha)|V\rangle +\sin(2\alpha)|H\rangle$ state, where $\alpha$ is the physical angle of
the half-wave plate. The $H$ component of this rotated state exits the interferometer, transmitted into mode $b$ with probability $p=\sin^{2}(2\alpha)$,
while the vertical component is reflected into mode $a$ with probability $1-p=\cos^{2}(2\alpha)$.

In turn, the $H$ photons of the input state propagate through the interferometer in the
clockwise direction and exit through the PBS into mode $a$, with their
polarization state unaltered.
In this way, we obtain $|H\rangle|a\rangle\rightarrow|H\rangle|a\rangle$
and $|V\rangle|a\rangle\rightarrow\sqrt{1-p}|V\rangle|a\rangle
+\sqrt{p}|H\rangle|b\rangle$, which is equivalent to map \eqref{eq:ADCmap} with
$p=\sin^{2}(2\alpha)$. A HWP oriented at $0^{\circ}$ (H0) is placed on the $H$ photons
path to compensate for the optical path difference. The relative path length of interferometer
is adjusted so that when H1 is oriented at $0^{\circ}$ the polarization
of the input state remains unaltered by the interferometer.

Photons in mode $a$ and $b$ are then measured by performing standard quantum
state tomography of the polarization state, using a quarter-wave plate, a
half-wave plate and a polarizing beam splitter, which recombines mode $b$ into
mode $a$ incoherently so that both modes can be detected with a single photon
detector. In this way, the environment is traced out and the original system is left affected 
by the noisy channel. With this purpose, $H$ polarized photons on mode $b$ are rotated by a HWP
oriented at $45^{\circ}$ (H2) just before passing through the PBS.

The phase damping channel can also be implemented by simply removing H2 from mode $b$. Since only horizontally polarized photons are routed into mode $b$ at the output of the PBS, this is equivalent to encode mode $b$ into the vertical polarization. The resulting map corresponds to a phase damping interaction described by equation \eqref{eq:PDCmap}, once the environment is traced out by incoherent combination of both spatial modes \cite{almeida2007environment,salles2008experimental}.

\begin{figure}[h!]
   \centering
   \includegraphics[width=0.45\textwidth]{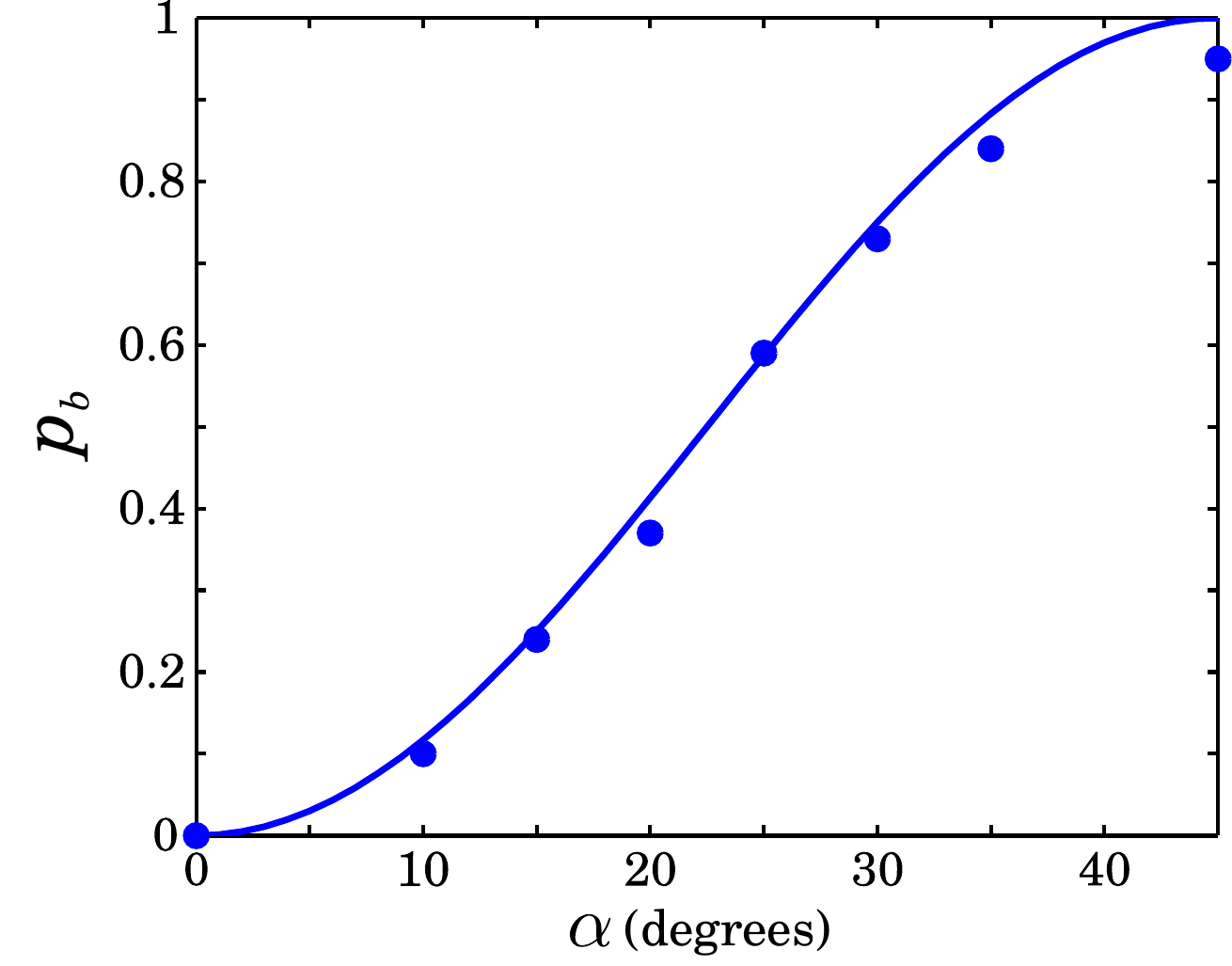}
 \caption{Measured values of $p_b$ as a function of the half-wave
 plate's angle $\alpha$ shown in dots. The solid line corresponds to the theoretical prediction for
 $p_b=\sin^{2}(2\alpha)$.}
 \label{fig:calibracionBOB}
 \end{figure}

The characterization and parametrization of this noisy environment was done as follows: for each value of the half waveplate angle $\alpha$, we measured and reconstructed the two-photon state, and we obtained the corresponding 
value of $p_b$ by finding the maximum fidelity  between the measured
density matrix (after a maximum likelihood optimization) and the
predicted density matrix. This matrix was calculated by numerically applying the amplitude damping map
to the available entangled state, measured for $p_a=p_b=0$. The measured values as a function of the
half-wave plate's angle $\alpha$ are shown in Fig. \ref{fig:calibracionBOB} in
blue dots and the theoretical prediction for $p_b=\sin^{2}(2\alpha)$ is shown in
solid blue line.

\emph{Local noise on Alice's qubit.} This polarization-based Sagnac interferometer is a useful tool for studying
quantum channels in a controlled manner. 
In order to investigate the behavior of the teleportation algorithm under local coupling to noisy environments, 
we also need to consider the effect of a damping channel on Alice's side. Using the spatial qubit to encode the environment is not possible, since this degree of freedom is already used to encode the input state. Furthermore, the Sagnac interferometer produces two outputs that are spatially separated, and implementing a noise stage equivalent to the above described would imply building two identical, matched Sagnac polarizing interferometers, one for each path output. The amplitude damping channel on Alice's side was therefore simulated as a weighed average of the measurements obtained with two different pure states acting as the quantum resource, which were produced with the sole inclusion of a half wave plate in the path of one of the photons from the entangled pair and a rotation of the half wave plate that sets the polarization of the pump beam. Eq. \eqref{eq:ADCalice} can be re-written as:
\begin{equation}
	 \rho'=\frac{1}{2}
  \left( \begin{matrix}
    1 & 0 & 0 & \sqrt{1-p_{a}}\\
    0 & 0 & 0 & 0 \\
    0 & 0 & 0 & 0 \\
    \sqrt{1-p_{a}} & 0 & 0 & (1-p_{a})
    \end{matrix} \right)+    \frac{p_a}{2}
  \left( \begin{matrix}
    0 & 0 & 0 & 0\\
    0 & 1 & 0 & 0 \\
    0 & 0 & 0 & 0 \\
    0 & 0 & 0 & 0
    \end{matrix} \right);
\end{equation}
which can be interpreted as 
\begin{equation}
\rho'=\left(1-\frac{p_a}{2}\right)\rho'_1+\frac{p_a}{2}\rho'_2,
\end{equation}
with
\begin{multline}
\rho_1' = \sin^2\phi\left | HH \right \rangle \left \langle HH\right |+\sin\phi\cos\phi\left | HH \right \rangle \left \langle VV\right |\\ 
+ \sin\phi\cos\phi\left | VV \right \rangle \left \langle HH\right |+\cos^2\phi\left | VV \right \rangle \left \langle VV\right |
\label{eq:rho1}
\end{multline}
using the identification $\sin\phi=1/\sqrt{2-p_a}$;
and $\rho'_2=|HV\rangle\langle HV|$.

The pure state $\rho'_1$ can be generated by simply rotating the angle $\theta$ of the
half-wave plate placed before the BBO crystals, used to control the polarization of the pump beam (taking $\phi=2\theta$). Using the standard nonlinear crystal arrangement for the generation of type-I entangled states \cite{kwiat1999ultrabright}, this waveplate (H3 in Fig. \ref{fig:arreglo}) transforms the horizontally polarized pump beam into a diagonally polarized beam when its fast axis is rotated 22.5$^{\circ}$ from the vertical position. This configuration leads to a value of $p_a=0$. Instead, rotating the waveplate 45$^{\circ}$ produces a vertically polarized pump beam on the BBO crystals and generates a state $|HH\rangle\langle HH|$, which is therefore equivalent to  set $p_a=1$.

The other pure state $\rho'_2$ can be generated by rotating H3 45$^{\circ}$, and adding another HWP at 45$^{\circ}$ on Bob's path to further transform the downconverted pair into the state $|HV\rangle\langle HV|$.
The amplitude damping channel acting on Alice is then simulated by averaging the results of the experiment using these two states with their respective statistical weights, $\left(1-\frac{p_a}{2}\right)$ and $\frac{p_a}{2}$. 

\begin{figure}[h!]
   \centering
   \includegraphics[width=0.47\textwidth]{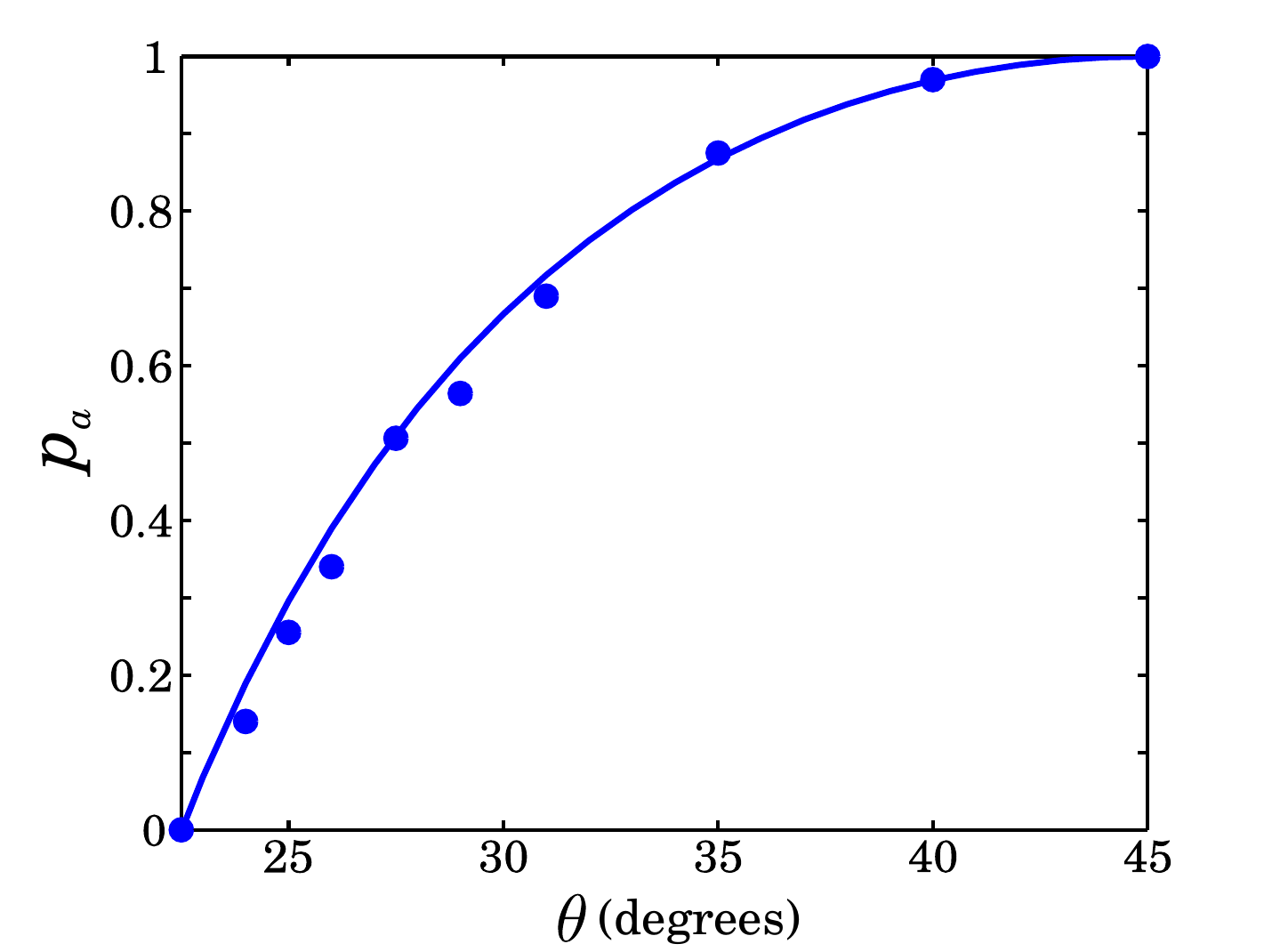}
 \caption{Measured values of $p_a$ as a function of the half-wave plate's
 angle $\theta$ shown in dots, together with the theoretical prediction for
 $p_a=2-\frac{1}{\sin^{2}(2\theta)}$ shown in solid line.}
 \label{fig:calibracionALICE}
 \end{figure}

In order to verify the parametrization for $p_a$ as a function of the experimentally accesible variable $\theta$, we calculated the density matrix $\rho_1'$ using different values of the waveplate angle. The corresponding coupling strength $p_a$ was obtained  by maximizing the fidelity between such density matrix and the density matrix calculated by applying the transformation described in Eq. \eqref{eq:rho1} to our original, experimentally limited entangled resource. Figure \ref{fig:calibracionALICE} shows the measured values of $p_a$ as a function of the pump's waveplate angle $\theta$ and the theoretical function $p_a=2-1/\sin^{2}(2\theta)$, valid for angles between 22.5$^{\circ}$ and 45$^{\circ}$.

Measurement of the fidelity of teleportation over different dissipative environments
was obtained as follows: for fixed values of $p_a$ and $p_b$, quantum process
tomography was performed for each of the possible outcomes of the teleportation
process, that is, the quantum state obtained by Bob conditioned to the
four possible results obtained by Alice. 
The value of the average process fidelity was calculated as shown in Eq. \eqref{eq:fidel_proba} for each condition of
$p_a$ and $p_b$.

\section{results and discussion}
\label{sec:results}

\begin{figure}[h!]
  \centering
  \includegraphics[width=0.45\textwidth]{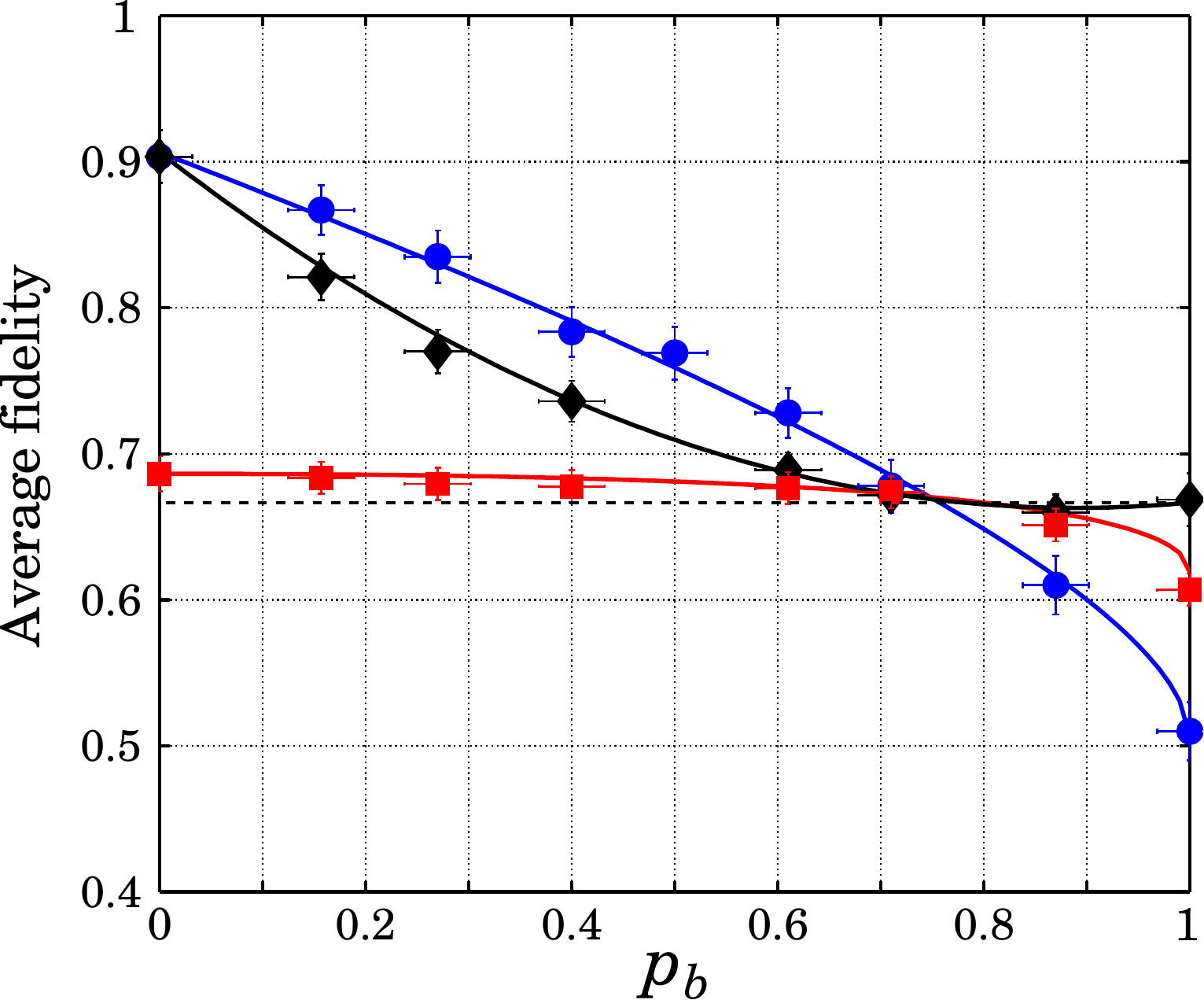}
    \caption{(Color online) Fidelity of teleportation for different
dissipative interactions via an amplitude damping channel on each path. The
measured data correspond to different values of $p_b$ on Bob's local environment
for three values of Alice's damping parameter: $p_a=0$ (blue circles), $p_a=0.7$
(red squares) and $p_a=p_b$ (black diamonds). The solid lines represent the
theoretical predictions obtained by applying the ADC map numerically to the
experimentally measured two-photon entangled initial state. The dashed line
corresponds to the classical limit of $F=2/3$.}
  \label{fig:fideliADC}
\end{figure}

\emph{Amplitude damping over both subsystems. }The action of local environments on the fidelity of quantum
teleportation was evaluated using a non-ideal polarization entangled state produced by SPDC 
and limited by experimental conditions. This state has a fidelity 
with the $|\Phi^+\rangle$ Bell state of $85\%$, as defined in \cite{gilchrist2005distance}. 
By applying the map \eqref{eq:ADCmap}
to a tomographic reconstruction of this entangled state, different theoretical curves can be
predicted for the fidelity of teleportation over local, bipartite, amplitude damping
channels. Figure \ref{fig:fideliADC} shows the predicted
theoretical curves and the experimentally measured fidelity for
different values of $p_b$, for different damping conditions on Alice's
environment ($p_a$). The dashed line corresponds to the classical limit of $F=2/3$. 
The vertical error bars were calculated by
Monte-Carlo simulation of experimental runs with the same Poissonian statistics.
Horizontal error bars were obtained from the standard deviation of the
measured values of $p_b$ for repeated measurements.

The measured points show good agreement with the theoretical prediction for this particular
noise channel. As expected, for a fixed value of $p_a$ the fidelity of quantum teleportation is a monotonically 
decreasing function of $p_b$ (blue and red points). The black points and curve correspond to the condition $p_a$=$p_b$. In section \ref{sec:dissipenv}, for the particular condition $p_a=2\sqrt{2}-2$, it was shown that an enhancement of the process fidelity can be obtained by increasing the amount of coupling with Bob's local environment to a value $p_b=p_a$. Even using a pure entangled resource, the expected enhancement is marginal. 
Using a tomographic reconstruction of our imperfect entangled pair, the point where our predicted curve for $p_a=0$
crosses the $F$=2/3 line was numerically found to be $p_b$=0.7568. Fixing $p_b$ at that value, and increasing the damping parameter on Alice's qubit up to $p_a$=0.48, we numerically obtain a maximum fidelity of $F\approx0.6725\gtrsim$2/3. 
The predicted enhancement is small and close to the experimental uncertainty. Nevertheless, repeated measurements were performed on these two points to obtain a mean value and standard deviation for the fidelity. For $p_b$ fixed at 0.73$\pm$0.05, we experimentally reproduce the classical limit of teleportation, $F$=0.667 for $p_a$=0. When $p_a$ is set to 0.47$\pm$0.05, the fidelity rises to $F$=0.675, with a standard deviation $\sigma$=0.005. This small increase on the teleportation fidelity becomes practically masked by the statistical error of our setup. Statistical fluctuations are indeed present in any physical implementation of the protocol, and due to the relatively small achievable enhancement, a statistical dispersion below 1\% is required for the effect to be noticeable.
Of more practical interest is the behavior reached for a heavy damping condition ($p_a$=0.7, red squares), where the fidelity of the quantum teleportation protocol becomes practically insensitive to noise on the other qubit, and yet it remains above the classical limit up to $p_b\approx$0.8.

\begin{figure}[h!]
  \centering
  \includegraphics[width=0.45\textwidth]{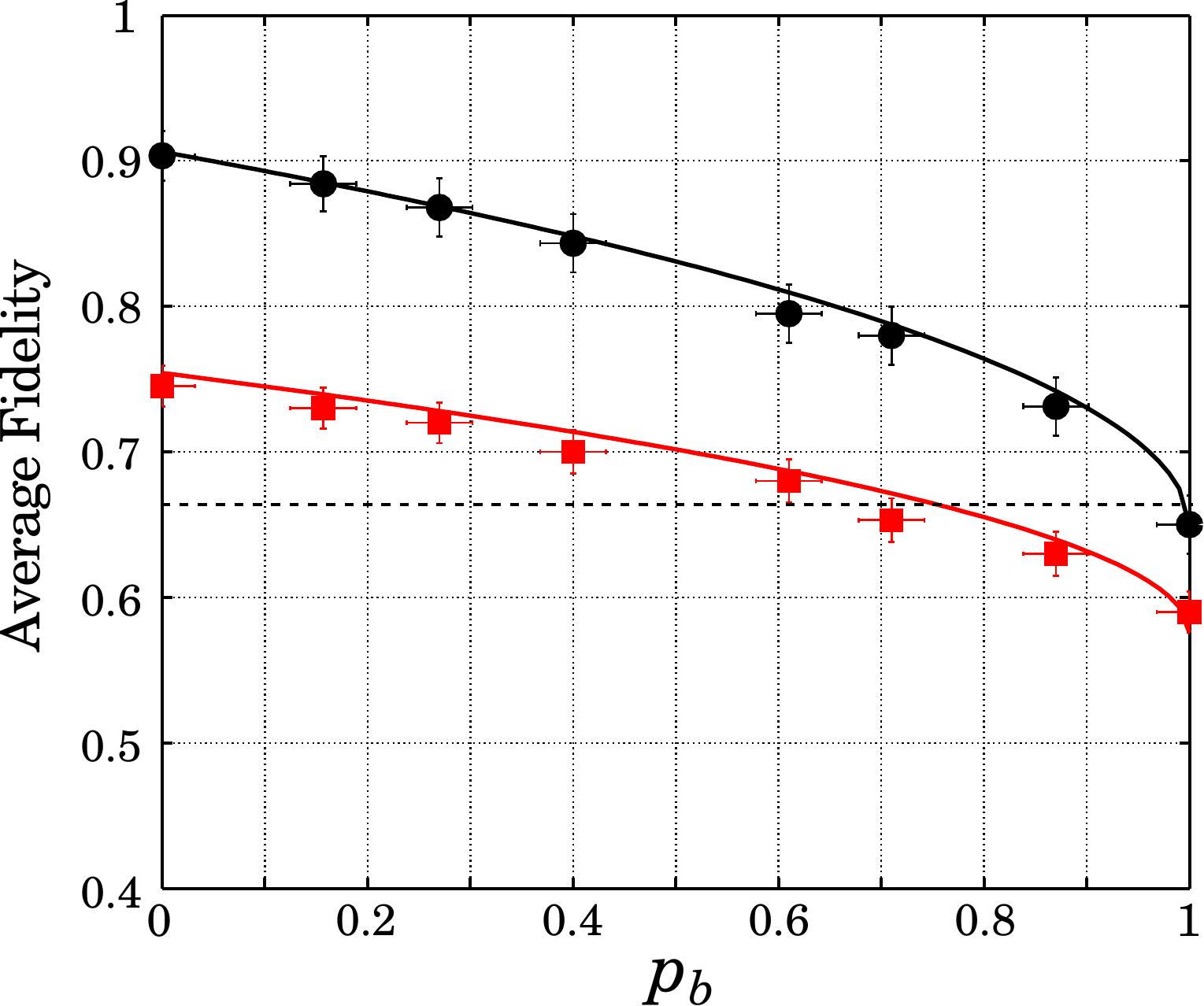}
    \caption{(Color online) Fidelity of teleportation for different
dissipative interactions via a phase damping channel on Bob's side and an
amplitude damping channel on Alice's side. The measured data correspond to
different values of $p_b$ on Bob's local environment for two values of Alice's
damping parameter: $p_a=0$ (black circles) and $p_a=0.5$ (red squares). The
solid lines represent the theoretical predictions obtained by applying the two
maps numerically to the experimentally measured two-photon entangled initial
state. The dashed line corresponds to the classical limit of $F=2/3$.}
  \label{fig:fideliPDC}
\end{figure}

\emph{Phase damping over one subsystem.} The protocol was repeated, now letting Bob's photon interact via a phase damping
channel, while Alice's photon still interacts via the amplitude damping channel.
Fig. \ref{fig:fideliPDC} shows the predicted theoretical curves and the
experimentally measured data of the fidelity for different values of $p_b$, for
two different amplitude damping parameters on Alice's environment: $p_a=0$
(black circles) and $p_a=0.5$ (red squares). The dashed line corresponds to
the classical limit of $F=2/3$. Vertical and horizontal error bars were obtained
as before. 

As expected, no enhancement can be obtained by adding noise to the system via a phase damping channel when the other photon is subjected to
an amplitude damping channel. Nevertheless, the action of the phase damping channel over one qubit (that is, for $p_a=0$), despite destroying the quantum coherence of the entangled pair, produces fidelities above the classical limit for almost all values of $p_b$, unlike applying an ADC over one qubit. 

\section{concluding remarks}
\label{sec:conclusions}
We have presented experimental results that investigate the influence of local environments on the fidelity of the teleportation protocol. Taking into account non-ideal, realistic entangled resources, we studied the interaction of the two qubits of the entangled pair with a local environment. By applying an ADC on both parties, different theoretical conditions were identified and measured experimentally on a photonic implementation of the teleportation algorithm. 
We obtained a good agreement with the theoretical model for dissipative channels. The predicted enhancement of the teleportation fidelity by an increase of  the coupling to a local environment is marginal, and in practice it might be easily masked by statistical noise. Regions of increased insensibility to amplitude damping noise could be observed in the experiment. 
The setup was also used to test the influence of a phase damping channel on Bob's side, using the same interferometer with a simple modification. The experimental results are again in good agreement with the predicted theoretical curves. 
This versatile setup allowed us to study different noisy channels in a repeatable manner, obtaining the general behavior of the action of local environments on the fidelity of the teleportation protocol despite having a limited entangled resource. The study of the influence of noise on particular protocols may lead to more robust, noise-insensitive implementations of quantum information processes.


The experiments were performed in the quantum optics laboratory at CITEDEF. LTK has a CONICET scholarship and MAL
is a CONICET fellow. The authors acknowledge fruitful discussions with
B.Taketani, J.P.Paz, A.Roncaglia.

\bibliographystyle{unsrtnat}
\bibliography{referencias}

\begin{thebibliography}{22}
\providecommand{\natexlab}[1]{#1}
\providecommand{\url}[1]{\texttt{#1}}
\expandafter\ifx\csname urlstyle\endcsname\relax
  \providecommand{\doi}[1]{doi: #1}\else
  \providecommand{\doi}{doi: \begingroup \urlstyle{rm}\Url}\fi

\bibitem[Bennett et~al.(1993)Bennett, Brassard, Cr{\'e}peau, Jozsa, Peres, and
  Wootters]{bennett1993teleporting}
Charles~H Bennett, Gilles Brassard, Claude Cr{\'e}peau, Richard Jozsa, Asher
  Peres, and William~K Wootters.
\newblock Teleporting an unknown quantum state via dual classical and
  einstein-podolsky-rosen channels.
\newblock \emph{Physical Review Letters}, 70\penalty0 (13):\penalty0 1895,
  1993.

\bibitem[Massar and Popescu(1995)]{massar1995optimal}
Serge Massar and Sandu Popescu.
\newblock Optimal extraction of information from finite quantum ensembles.
\newblock \emph{Physical review letters}, 74\penalty0 (8):\penalty0 1259, 1995.

\bibitem[Popescu(1994)]{popescu1994bell}
Sandu Popescu.
\newblock Bell’s inequalities versus teleportation: What is nonlocality?
\newblock \emph{Physical review letters}, 72\penalty0 (6):\penalty0 797, 1994.

\bibitem[Bennett et~al.(1996{\natexlab{a}})Bennett, Brassard, Popescu,
  Schumacher, Smolin, and Wootters]{bennett1996purification}
Charles~H Bennett, Gilles Brassard, Sandu Popescu, Benjamin Schumacher, John~A
  Smolin, and William~K Wootters.
\newblock Purification of noisy entanglement and faithful teleportation via
  noisy channels.
\newblock \emph{Physical Review Letters}, 76\penalty0 (5):\penalty0 722,
  1996{\natexlab{a}}.

\bibitem[Horodecki et~al.(1996)Horodecki, Horodecki, and
  Horodecki]{horodecki1996teleportation}
Ryszard Horodecki, Micha{\l} Horodecki, and Pawe{\l} Horodecki.
\newblock Teleportation, bell's inequalities and inseparability.
\newblock \emph{Physics Letters A}, 222\penalty0 (1):\penalty0 21--25, 1996.

\bibitem[Bennett et~al.(1996{\natexlab{b}})Bennett, DiVincenzo, Smolin, and
  Wootters]{bennett1996mixed}
Charles~H Bennett, David~P DiVincenzo, John~A Smolin, and William~K Wootters.
\newblock Mixed-state entanglement and quantum error correction.
\newblock \emph{Physical Review A}, 54\penalty0 (5):\penalty0 3824,
  1996{\natexlab{b}}.

\bibitem[Horodecki et~al.(1999)Horodecki, Horodecki, and
  Horodecki]{horodecki1999general}
Micha{\l} Horodecki, Pawe{\l} Horodecki, and Ryszard Horodecki.
\newblock General teleportation channel, singlet fraction, and
  quasidistillation.
\newblock \emph{Physical Review A}, 60\penalty0 (3):\penalty0 1888, 1999.

\bibitem[Badziag et~al.(2000)Badziag, Horodecki, Horodecki, and
  Horodecki]{badziag2000local}
Piotr Badziag, Micha{\l} Horodecki, Pawe{\l} Horodecki, and Ryszard Horodecki.
\newblock Local environment can enhance fidelity of quantum teleportation.
\newblock \emph{Physical Review A}, 62\penalty0 (1):\penalty0 012311, 2000.

\bibitem[Bandyopadhyay(2002)]{bandyopadhyay2002origin}
Somshubhro Bandyopadhyay.
\newblock Origin of noisy states whose teleportation fidelity can be enhanced
  through dissipation.
\newblock \emph{Physical Review A}, 65\penalty0 (2):\penalty0 022302, 2002.

\bibitem[Yeo(2008)]{yeo2008local}
Ye~Yeo.
\newblock Local noise can enhance two-qubit teleportation.
\newblock \emph{Physical Review A}, 78\penalty0 (2):\penalty0 022334, 2008.

\bibitem[Taketani et~al.(2012)Taketani, de~Melo, and
  de~Matos~Filho]{taketani2012optimal}
Bruno~G Taketani, Fernando de~Melo, and RL~de~Matos~Filho.
\newblock Optimal teleportation with a noisy source.
\newblock \emph{Physical Review A}, 85\penalty0 (2):\penalty0 020301, 2012.

\bibitem[Nielsen and Chuang(2010)]{nielsen2010quantum}
Michael~A Nielsen and Isaac~L Chuang.
\newblock \emph{Quantum computation and quantum information}.
\newblock Cambridge university press, 2010.

\bibitem[Kraus(1983)]{kraus1983states}
Karl Kraus.
\newblock \emph{States, effects and operations}.
\newblock Springer, 1983.

\bibitem[Grondalski et~al.(2002)Grondalski, Etlinger, and
  James]{grondalski2002fully}
J~Grondalski, DM~Etlinger, and DFV James.
\newblock The fully entangled fraction as an inclusive measure of entanglement
  applications.
\newblock \emph{Physics Letters A}, 300\penalty0 (6):\penalty0 573--580, 2002.

\bibitem[Knoll et~al.(2014)Knoll, Schmiegelow, and Larotonda]{knoll2014remote}
Laura~T Knoll, Christian~T Schmiegelow, and Miguel~A Larotonda.
\newblock Remote state preparation of a photonic quantum state via quantum
  teleportation.
\newblock \emph{Applied Physics B}, 115\penalty0 (4):\penalty0 541--546, 2014.

\bibitem[Kwiat et~al.(1999)Kwiat, Waks, White, Appelbaum, and
  Eberhard]{kwiat1999ultrabright}
Paul~G Kwiat, Edo Waks, Andrew~G White, Ian Appelbaum, and Philippe~H Eberhard.
\newblock Ultrabright source of polarization-entangled photons.
\newblock \emph{Physical Review A}, 60\penalty0 (2):\penalty0 R773, 1999.

\bibitem[James et~al.(2001)James, Kwiat, Munro, and
  White]{james2001measurement}
Daniel~FV James, Paul~G Kwiat, William~J Munro, and Andrew~G White.
\newblock Measurement of qubits.
\newblock \emph{Physical Review A}, 64\penalty0 (5):\penalty0 052312, 2001.

\bibitem[Renes et~al.(2004)Renes, Blume-Kohout, Scott, and
  Caves]{renes2004symmetric}
Joseph~M Renes, Robin Blume-Kohout, Andrew~J Scott, and Carlton~M Caves.
\newblock Symmetric informationally complete quantum measurements.
\newblock \emph{Journal of Mathematical Physics}, 45:\penalty0 2171, 2004.

\bibitem[Bendersky et~al.(2009)Bendersky, Pastawski, and
  Paz]{bendersky2009selective}
Ariel Bendersky, Fernando Pastawski, and Juan~Pablo Paz.
\newblock Selective and efficient quantum process tomography.
\newblock \emph{Physical Review A}, 80\penalty0 (3):\penalty0 032116, 2009.

\bibitem[Almeida et~al.(2007)Almeida, de~Melo, Hor-Meyll, Salles, Walborn,
  Ribeiro, and Davidovich]{almeida2007environment}
MP~Almeida, Fernando de~Melo, Malena Hor-Meyll, Alejo Salles, SP~Walborn,
  PH~Souto Ribeiro, and Luiz Davidovich.
\newblock Environment-induced sudden death of entanglement.
\newblock \emph{Science}, 316\penalty0 (5824):\penalty0 579--582, 2007.

\bibitem[Salles et~al.(2008)Salles, de~Melo, Almeida, Hor-Meyll, Walborn,
  Ribeiro, and Davidovich]{salles2008experimental}
Alejo Salles, Fernando de~Melo, MP~Almeida, Malena Hor-Meyll, SP~Walborn,
  PH~Souto Ribeiro, and Luiz Davidovich.
\newblock Experimental investigation of the dynamics of entanglement: Sudden
  death, complementarity, and continuous monitoring of the environment.
\newblock \emph{Physical Review A}, 78\penalty0 (2):\penalty0 022322, 2008.

\bibitem[Gilchrist et~al.(2005)Gilchrist, Langford, and
  Nielsen]{gilchrist2005distance}
Alexei Gilchrist, Nathan~K Langford, and Michael~A Nielsen.
\newblock Distance measures to compare real and ideal quantum processes.
\newblock \emph{Physical Review A}, 71\penalty0 (6):\penalty0 062310, 2005.

\end{thebibliography}

\end{document}